\documentclass[letterpaper, 11 pt, peerreviewca]{ieeeconf}

\usepackage{courier}
\usepackage{hyperref}
\usepackage{graphicx,subfigure}
\usepackage{epstopdf}
\usepackage{tikz}
\usepackage{amsmath}
\usepackage{multirow}
\usepackage{amsfonts}
\usepackage{float}
\usepackage{amssymb}
\usepackage{graphicx}
\usepackage{flushend}
\usepackage[ruled]{algorithm}
\usepackage{algpseudocode}
\newtheorem{theorem}{Theorem}

\newtheorem{corollary}{Corollary}
\newtheorem{definition}{Definition}
\newtheorem{remark}{Remark}

\makeatletter

\newcommand{\Rmnum}[1]{\expandafter\@slowromancap\romannumeral #1@}

\makeatother

\begin{document}
\title{\LARGE Visual Navigation with a 2-pixel Camera---Possibilities and Limitations}
\author{John Baillieul and Feiyang Kang}
\maketitle
\let\thefootnote\relax\footnotetext{\noindent
\hspace{-0.1in}\hrulefill
\hspace{0.8in}\\John Baillieul is with the Departments of Mechanical Engineering, Electrical and Computer Engineering, and the Division of Systems Engineering at Boston University, Boston, MA 02115. He may be reached at {\tt johnb@bu.edu}. \newline Support from various sources including the Office of Naval Research grants N00014-10-1-0952, N00014-17-1-2075, and N00014-19-1-2571 is gratefully acknowledged.\\
A condensed version of this paper has appeared in Proceedings of IFAC 2020, Virtual World Congress, Berlin, July 13-17,2020,} 

\begin{abstract}
\noindent Borrowing terminology from fluid mechanics, the concepts of {\em Eulerian} and {\em Lagrangian optical flow sensing} are introduced.  Eulerian optical flow sensing assumes that each photoreceptor in the camera or eye can instantaneously detect feature image points and their velocities on the retina.  If this assumption is satisfied, even a two pixel imaging system can provide a moving agent with information about its movement along a corridor that is sufficiently precise as to be used as a robustly reliable steering signal.  Implementing Eulerian optical flow sensing poses significant challenges, however.  Lagrangian optical flow, on the other hand, tracks feature image points as they move on the retina.  This form of visual sensing is the basis for many standard computer vision implementations, including Lukas-Kanade and Horn-Schunck.  Lagrangian optical flow has its own challenges, not least of which is that it is badly confounded by rotational components of motion.  Combined steering and sensing strategies for mitigating the effects of rotational motions are considered.
\end{abstract}
\begin{flushleft} {\bf Keywords:} Eulerian optical flow, Lagrangian optical flow.
\end{flushleft}


 \section{Introduction}

Successful emulation of the way animals use vision to guide their movement requires accurate and computationally tractable models of the way in which they interpret multiple visual cues and they ways in which they validate information streams coming from such cues using other sensory modalities.  Depth cues coming from both binocular vision as well as from optical flow are central to animal navigation, and recent advances in light-weight high-speed computer hardware have made it possible to perform the necessary computations using established sparse optical flow algorithms such as Lucas-Kanade, \cite{Lucas} and Horn-Schunck, \cite{Horn}.  Challenges remain, however, in implementing bio-inspired robot control based on optical flow due to the complexity of extracting reliable steering signals from a moving camera.  See, e.g.\  \cite{Marquez}, \cite{Corvese}, and \cite{Seebacher}.

Factors that confound perception based on visual cues include 
\begin{itemize}
\item {\em Depth discontinuities} associated with obstacle boundaries that may be difficult to distinguish from noise in optical flow,
\item Moving objects in the field of view that produce localized optical flow that is inconsistent with the optical flow that is generated by self-motion,
\item Ephemeral persistence of features within the field of view (FoV),
\item Flow indeterminacy due to very sparse optical sensor data in part of the FoV,
\item Rotational movement of the optical sensor relative to the features being observed.
\end{itemize}
Moving objects, of course, produce depth (and hence optical flow) discontinuities in the same way that stationary objects do, but they are further confounding when they generate a spurious local focus of expansion, \cite{Layton}.  In order for optical flow to generate reliable guidance for movement, frame-to-frame persistence of the images of features on which the flow is based is essential.  Such persistence can be highly elusive for a number of reasons including a lack of framewise invariance of a feature's descriptor or the feature moving too rapidly out of the field of view.  Feature descriptor errors can occur because of changes in lighting or occultation due to loss of line-of-sight visibility associated with movement past an obstacle.  If the camera passes too close to the images of features to be used in computing optical flow will disappear from the image plane of the camera too quickly to be of use in determining flow vectors.  (Think of the difference between the speeds with which images of objects nearby and far away enter and then disappear from the field of view depending on their distance from a moving car.)  We believe that difficulty obtaining reliable optical flow from features that are too close to a moving agent may explain field observations of flying bats whose flight paths were parallel to but displace from the edge of a stand of tree by about three meters, \cite{SciRepts}. 

The fourth and fifth confounding items may be the most significant.  Featureless regions in the field of view associated with, say, a perfectly monochromatic wall will give rise to sparsity of image features that render optical flow calculations unreliable.  Even in environments where features are abundant and have images that persist frame-to-frame, there will generally be a small region surrounding the focus of expansion (FoE) where flow vectors are noisy and provide little reliable information about the motion.  Optical flows around the FoE are particularly badly distorted by a rotational component along a path, but as we show below, the flow field in all parts of the image are affected.  

\section{Idealized optical flow}

Over many decades, a great deal of research has been devoted to measuring and mitigating the effects of rotational disturbances \cite{Royden}  While a number of mechanism for determining heading based on optical flow have been proposed---see, for example, \cite{Layton}---satisfactory solutions to reliable navigation have remained elusive.  It is against this backdrop that we hope to obtain insights from the idealized models proposed in what follows.   Specifically, we shall consider optical flow based steering laws using the perceived quantity {\em time-to-transit} along the lines introduced in \cite{Sebesta}, \cite{Kong}, \cite{Seebacher}, \cite{Corvese}, and \cite{CDC19}.

While there are relatively few optical sensing technologies that could realize our idealized model (See, however, \cite{Gupta}.), the aim is to derive insights into the way movement affects visual sensing.  The goal is to understand geometric aspects of the factors that confound visual sensing by moving agents in the highly idealized setting of perfect instantaneous perception of single features.  As explained in \cite{CDC19}, the main purpose of our use of these idealized models is to build intuition and suggest strategies for extracting steering signals from sparse optical flow recovered from imagers on moving robots.  The control law of \cite{CDC19} was shown to robustly steer a unicycle robot along the center line of a corridor under the assumption that an infinitely dense set of features were available and continuously detected by the photo receptors.  For this idealized setting, the steering laws are unaffected by moderate rotational perturbations along the path.   When the assumptions underlying the idealization are relaxed,  and features are modeled as discrete points that are sparsely distributed in the field of view, the effects of rotational motion components are severe.  Over many decades, a great deal of research has been devoted to measuring and mitigating the effects of rotational disturbances---see, for example, \cite{Royden} and \cite{Marquez}.   While valuable insights have been gleaned, satisfactory solutions to reliable navigation have remained elusive.  It is against this backdrop that we hope to obtain insights from the idealized models proposed below.   Several implementations with laboratory robots have been developed to examine the principles on actual hardware, but this work will be described elsewhere.

\begin{figure}[h]
\begin{center}
\includegraphics[scale=0.65]{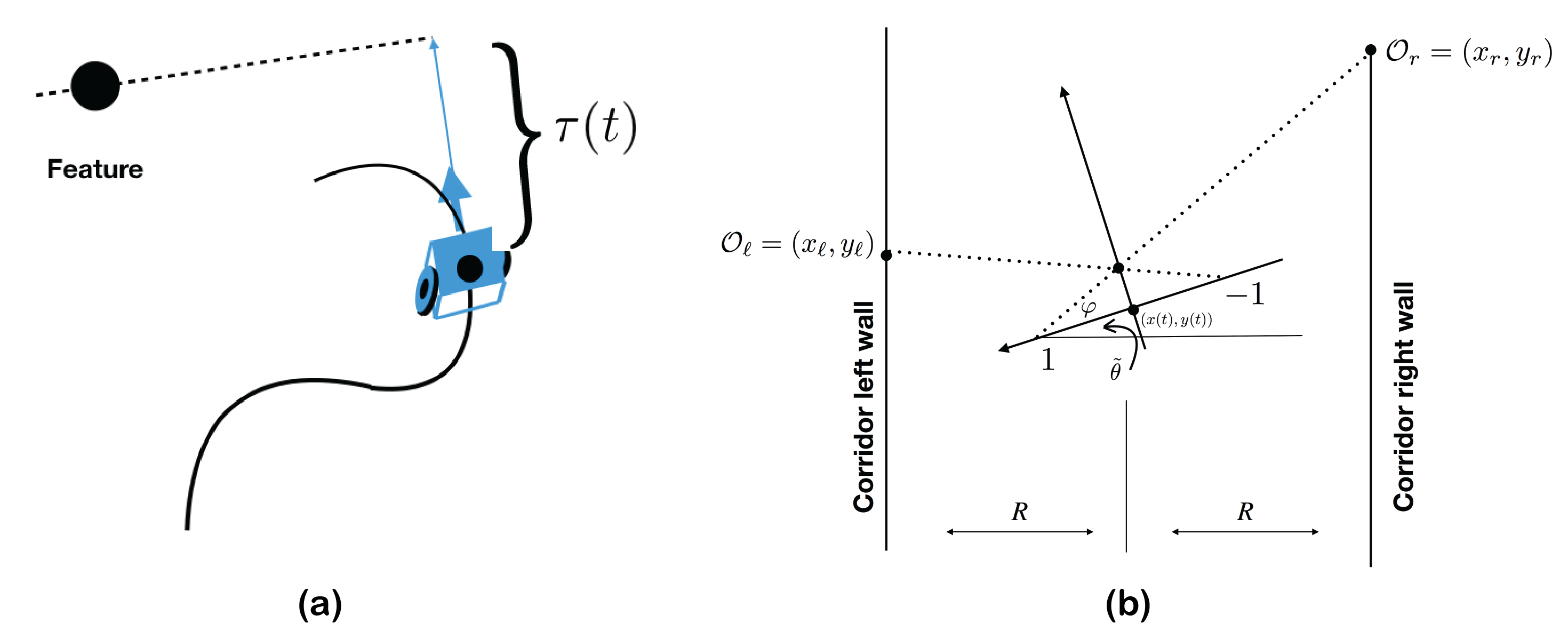}
\end{center}
\caption{Vehicle with kinematics (\ref{eq:jb:BasicVehicle}).  Here, the stylized image plane is compressed to one dimension and represented by a line segment on the vehicle $y$-axis. The vehicle $x$-axis is the direction of travel, $\tan\varphi=f$, the pinhole camera focal length, and $\theta=\tilde\theta+\frac{\pi}2$.  {\bf (a)} Time-to-transit, $\tau$, is available on the image plane as $\tau=r/\dot r$, with $r$ the signed distance from the body frame origin (\cite{CDC19}). {\bf (b)} The spatial features ${\cal O}_r$ and ${\cal O}_{\ell}$ are registered at $\pm 1$ respectively on the image plane ($y$-axis).  With $x_r=R,\ x_{\ell}=-R$, simple geometry shows that $y_{\ell}=y+f \sin (\theta )+\frac{(R+x+f \cos (\theta )) (\cos (\theta )+f \sin (\theta ))}{\sin(\theta )-f \cos (\theta )}$ and $y_r= y+f \sin (\theta )+\frac{(R-x-f \cos (\theta )) (f \sin (\theta )-\cos (\theta ))}{f
   \cos (\theta )+\sin (\theta )}$.}
\label{fig:RobotFigs}
\end{figure}

{\bf Time-to-Transit as a perceived quantity}  In \cite{Sebesta},\cite{Kong},\cite{Seebacher},\cite{Corvese}, and \cite{CDC19}, optical flow based steering laws using the perceived quantity {\em time-to-transit} have been proposed.  Recall that if a vehicle is moving in a straight line and a feature point lies somewhere ahead of the vehicle---possibly to the left or right in the environment, we can draw a plane that passes through the feature point and is also perpendicular to the line of travel.  The {\em time-to-transit} the feature is defined as the time it will take the vehicle to follow the straight line until it reaches the plane---as illustrated in 2D in Fig.\ \ref{fig:RobotFigs}.  As noted in the references above, under the assumption  that the vehicle follows a straight line at a constant speed, the time-to-transit can be determined from the movement of feature image points on the image plane of a forward looking camera.  Let $r$ denote the radial distance from a feature's image point on the image
 plane to the point where the optical axis intersects the image plane.  Then the {\em time-to-transit} the feature point, denoted by, $\tau$, is given by $r/\dot r$.  The reader is referred to Fig.\ 1(a) and \cite{CDC19} for details.

\begin{remark} \rm
Because $\tau$ can be determined by image motions in the image plane (or retina), it is not surprising that there is strong experimental evidence that it is something that animals can perceive.  It must be noted, however, that $\tau$ as perceived as $r/\dot r$ is only equal to the actual time-to-transit when the velocity in the direction fo the feature is constant.  This constrains our use of $\tau$ as a proxy for feature distance in our steering laws.
\end{remark}

The idealized steering law described in \cite{CDC19} is based on two assumptions.  First, it is assumed that at each of two preselected photo receptors, a unique feature in the environment is registered.  Second, it is assumed that under the current heading of the vehicle, the $\tau$-values for the features registered at the photo receptors are instantaneously determined.  Under these assumptions, if the photoreceptors are symmetrically located on opposite sides of the optical axis, as in Fig.\ 1(b), a unicycle vehicle with kinematics

\begin{equation}
\left(\begin{array}{c}
\dot x \\
\dot y \\
\dot\theta\end{array}\right) = \left(\begin{array}{c}
v\cos\theta \\
v\sin\theta \\
u\end{array}\right),
\label{eq:jb:BasicVehicle}
\end{equation}
where $v$ is the forward speed in the direction of the body-frame $x$-axis, and $u$ is the turning rate,  can be stably steered along the centerline of a corridor by balancing the $\tau$-values of features ${\cal O}_{\ell}$ and ${\cal O}_r$ on opposite corridor walls.  From \cite{CDC19}, we have

\begin{theorem} \rm
Consider a mobile camera moving along an infinitely long corridor with every point along both walls being a detectable feature that determines an accurate value of $\tau$.   Suppose the corridor has width $2R$ as depicted in Fig.~\ref{fig:RobotFigs}{\bf (b)}.   Let $\tau_r=\tau({\cal O}_r)$ and $\tau_{\ell}=\tau({\cal O}_{\ell})$ be the respective times to transit the two feature points whose images appear at points equidistant on either side of the optical axis (at $\pm 1$).  Then for any gain $k>0$ there is an open neighborhood $U$ of $(x,\theta)=(0,\frac{\pi}{2})$, $U\subset\{(x,\theta)\; :\; -R<x<R;\; \varphi<\theta<\pi-\varphi\}$ such that for all initial conditions $(x_0,y_0,\theta_0)$ with $(x_0,\theta_0)\in U$, the steering law
\begin{equation}
u(t)=k(\tau_{\ell}-\tau_r)
\label{eq:jb:tau-balance}
\end{equation}
asymptotically guides the vehicle with kinematics (\ref{eq:jb:BasicVehicle}) onto the center line between the corridor walls.
\label{th:jb:one}
\end{theorem}
\begin{flushright} $\blacksquare$ \end{flushright}

This is proved in \cite{CDC19}.  It provides a remarkably robust steering law that works well even when some of the stated assumptions are significantly relaxed.  The requirement that the photoreceptors be symmetrically located on opposite sides of the optical axis is not needed, for instance, and we have the following.

\medskip 

\begin{corollary}\rm 
Consider a mobile camera moving along a corridor as in Theorem 1, and let $\tau_r==\tau({\cal O}_r)$ and $\tau_{\ell}=\tau({\cal O}_{\ell})$ be the respective times to transit  two feature points whose images register on opposite sides of the optical axis at $-\delta$ and $\epsilon$ respectively.   Then for any gain $k>0$ there is an open neighborhood $U\subset\{(x,\theta)\; :\; -R<x<R;\; \varphi<\theta<\pi-\varphi\}$ such that for all initial conditions $(x_0,y_0,\theta_0)$ with $(x_0,\theta_0)\in U$, the steering law
\begin{equation}
u(t)=k(\tau_{\ell}-\tau_r)
\end{equation}
asymptotically guides the vehicle onto a line parallel to the corridor walls with the asymptotic limit $(x(t),\theta(t)) \ \to  \ \Large\left(R(\delta-\epsilon)/(\delta+\epsilon),\pi/2\Large\right)$.
\end{corollary}
\begin{proof}
The proof is essentially the same as for Theorem 1, except the geometry is changed such that the coordinates of the feature points are
\[
{\cal O}_{\ell}=\left(
\begin{array}{c}
 -R \\
 y+f \sin (\theta )+\frac{(R+x+f \cos (\theta )) (\delta\cos (\theta )+f \sin (\theta ))}{\delta\sin
   (\theta )-f \cos (\theta )} \\
\end{array}
\right)
\]
and
\[
{\cal O}_r=\left(
\begin{array}{c}
 R \\
 y+f \sin (\theta )+\frac{(R-x-f \cos (\theta )) (f \sin (\theta )-\epsilon\cos (\theta ))}{f \cos (\theta )+\epsilon\sin (\theta )} \\
\end{array}
\right).
\]
\end{proof}

\medskip 
\begin{remark}\rm
Clearly if one photoreceptor is significantly closer that the other to the optical axis, the steady-state motion will be closer to the wall opposite to the photoreceptor.
\end{remark}

\medskip

\begin{corollary}
Consider a mobile camera as in Corollary 1 with the same $\tau_{\ell}=\tau({\cal O}_{\ell}),\tau_r=\tau({\cal O}_r)$ and photoreceptors at the same $(-\delta, \epsilon)$ locations in the body frame.  For any gain $k>0$, there is an open neighborhood $\subset\{(x,\theta)\; :\; -R<x<R;\; \varphi<\theta<\pi-\varphi\}$ such that for all initial conditions $(x_0,y_0,\theta_0)$ with $(x_0,\theta_0)\in U$, the steering law
\begin{equation}
u(t)=k(\delta\tau_{\ell}-\epsilon\tau_r)
\label{eq:jb:weighted-tau-bal}
\end{equation}
asymptotically guides the vehicle onto a line parallel to the corridor walls with the asymptotic limit $(x(t),\theta(t)) \ \to  \ \Large\left((\epsilon-\delta)/2,\pi/2\Large\right)$.
\end{corollary}
\begin{flushright} $\blacksquare$ \end{flushright}

\section{Perceptual aliasing and quantization}

In the steering laws proposed in the last section, time-to-transit ($\tau$) has been taken as a proxy for depth or distance.  One possible approach to implementing such laws would be use a time-of-flight distance measuring device such as a LIDAR.  Such devices satisfy the assumptions of being able to perfectly and instantaneously sense the distance to features that are densely arrayed along corridor walls.  It is doubtful that general purpose video cameras can meet our assumptions of perfect sensing of infinitely dense arrays of visual features.  As noted in the introduction, rotational components of movement are confounding, and unless a rotational component  is infinitesimally small, it will cause large movements of image points in the image plane.  These will invalidate estimates of tau values based on movements of feature image points on the image plane.

To understand the problem and its possible solution, we examine the difference between a purely geometric definition of $\tau$ and the value of $\tau$ that is perceived by means of the movement of image points in the image plane.

\begin{definition}
Consider a feature point with coordinates $(x_f,y_f)$ and a vehicle whose configuration evolves according to (\ref{eq:jb:BasicVehicle}).  Given the current configuration $(x(t),y(t),\theta(t))$  and speed $v(t)$, the {\em geometric time-to-transit} ({\em geometric} $\tau$) is the time it would take the vehicle with its current speed $v(t)$ and heading $\theta(t)$ held constant  to cross a line intersecting the feature and perpendicular to the current heading.  
\end{definition}

\begin{remark}
Geometric tau is illustrated in Fig.\ \ref{fig:RobotFigs} and is can be shown to be given by the formula
\begin{equation}
\tau(t) = \frac{\cos\theta(t)(x_f-x(t))+\sin(t)(y_f-y(t))}{v}.
\label{eq:jb:GeomTau}
\end{equation}
\end{remark}
\smallskip

\begin{figure}[h]
\begin{center}
\includegraphics[scale=0.35]{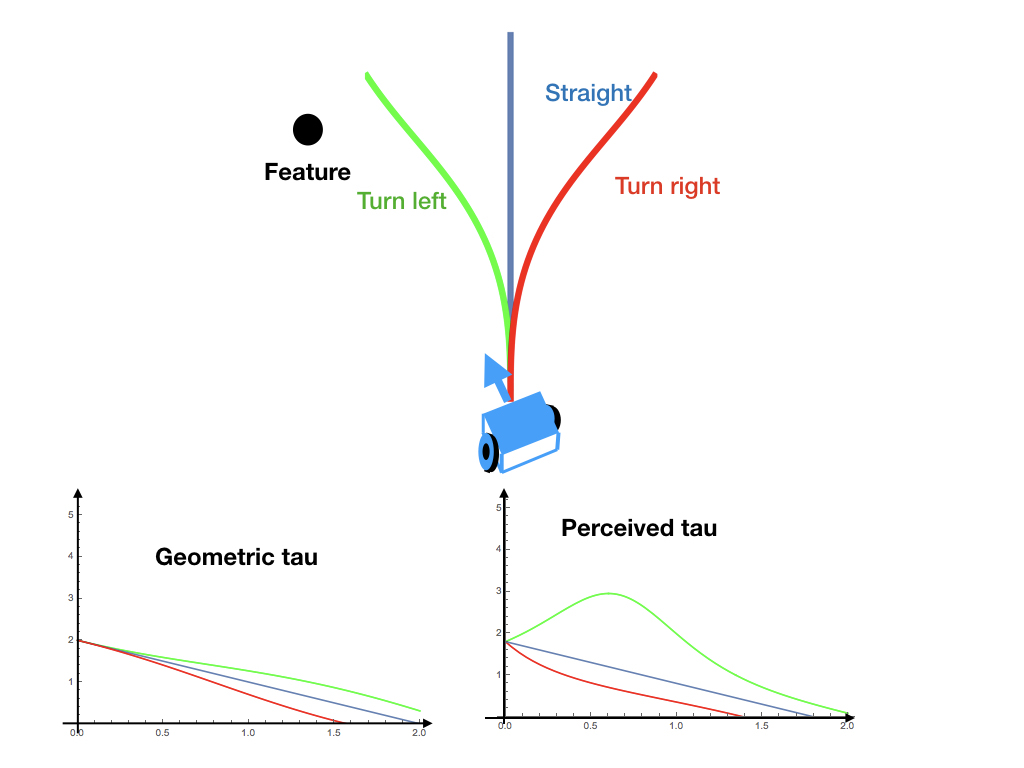}
\end{center}
\caption{When time-to-transit, $\tau$ is computed using pixel motions on the retina, values deviate from the corresponding geometric quantities.  Here, the vehicle turns slightly toward the feature as it rises toward the top of the figure along the green path.  While the geometry of this motion clearly increases $\tau$ relative to what it would have been traveling at the same speed along the straight (blue) path, the perceived value increases considerably more.}
\label{fig:jb:false-tau}
\end{figure}

\noindent It is easy to see that the value of geometric tau given by (\ref{eq:jb:GeomTau}) is maximized if the vehicle is headed directly toward the feature point, and it is zero if the heading direction is perpendicular to the vector from the current vehicle position $(x(t),y(t))$ to the featue point $(x_f,y_f)$.  If the vehicle is moving along a curved path, as illustrated in Fig.\ \ref{fig:RobotFigs}(a), this geometric value of tau will increase if the vehicle is turning toward the feature point and decrease if the vehicle is turning away.  If the values of $\tau$ are computed by means of the movement of image points on the image plane for the same motions, the values of $\tau$ as perceived in the image plane correspondingly increase or decrease, but they are significantly exaggerated as illustrated in Fig.\ \ref{fig:jb:false-tau}.

Time-to-transit os a proxy for distance, and {\em geometric tau} is taken to be ground truth in visual estimation of distance.  In light of the above remark and the essential assumption that both speed and heading \underbar{must} be constant in order for tau to be accurately perceived by the movement of image points on the retina or image plane, we must carefully determine the conditions under which perceived values of tau can be used for steering.  Thus it becomes important to examine the visual processes by which animals and machine simulacra perceive tau.  For the perceptual mechanisms that satisfy the assumptions of Theorem 1 and its corollaries, no single feature point has any significance.  The values of tau are instantaneously perceived, utilized in the steering law, and discarded in favor of a continuously updated stream of new values.  Borrowing terminology from fluid mechanics, we shall call optical flow tau-based perception in which the tau values are instantaneously sensed at a finite set of body-fixed photoreceptors {\em Eulerian optical flow sensing}.  The assumptions and validity of Theorem 1 depend on flow sensing being Eulerian.

By contrast, we shall call optical flow sensing in which discrete environmental feature points give rise to continuously moving image points on the image plane or retina {\em Lagrangian optical flow sensing}.  To the extent that animals, humans, and robots can implement Eulerian optical flow and tau sensing, they are able to employ the steering laws (\ref{eq:jb:tau-balance}) - (\ref{eq:jb:weighted-tau-bal}).  As mentioned above, depth cameras such as LIDARs can be used to implement Eulerian sensing.  Cameras with standard optics can approximately implement Eulerian flow sensing by using a restricted form of Lagrangian flow sensing under stringent conditions that include dense arrays of visible feature points, high densities of photoreceptors, and very modest rotational components of the motion.  

With less restrictive assumptions needed on movement, a hybrid form of Lagrangian tau-based flow sensing in which motions are segmented into sequences of {\em sense - perceive - act} components can be used to extract reliable steering signals.  The key is to produce a splined interpolation of the desired path by appropriately alternating between short straight path segments along which tau values for feature keypoints can be computed and curved segments whose constant curvatures are prescribed by tau values computed on the preceding straight segment.  The following provides part of the rationale for this approach.

\medskip

\begin{theorem}
Consider the planar vehicle (\ref{eq:jb:BasicVehicle}) for which the steering law is of the sample-and-hold type:
\begin{equation}
u(t)=k[\tau_{\ell}(x(t_i),\theta(t_i))-\tau_r(x(t_i),\theta(t_i))], \ \ t_i\le t<t_{i+1},
\label{eq:jb:sampled}
\end{equation}
where the sampling instants $t_0<t_1<\dots$ are uniformly spaced with $t_{i+1}-t_i = h>0$.  Then for any sufficiently small sampling interval $h>0$, there is a range of values of the gain $0<k<k_{crit}$ such that the sampled control law (\ref{eq:jb:sampled}) asymptotically guides the vehicle with kinematics (\ref{eq:jb:BasicVehicle}) onto the center line between the corridor walls.
\label{thm:jb:sampled}
\end{theorem}

\begin{proof}
As in the previous theorem, we assume that the forward speed is constant ($v=1$).  We also assume a normalization of scales such that $f=1$.  It is again convenient to consider the angular coordinate $\phi=\theta-\pi/2$.  In terms of this, we have
\[
\dot\phi = k[\tau_{\ell}(x(t_i),\phi(t_i)+\pi/{2})-\tau_r(x(t_i),\phi(t_i)+{\pi}/{2})]
\]
on the interval $t_i\le t < t_{i+1}$.  Given the explicit formulas for $\tau_{\ell}$ and $\tau_r$, and given that the right hand side of the above differential equation is constant, we have the following discrete time evolution
\begin{equation}
\phi(t_{i+1})=\phi(t_i) 
+ h k\frac{ 2 \sin \phi (t_1) (R+\cos \phi(t_i ))-2 x(t_i) \cos \phi(t_i )}{\sin ^2\phi (t_i)-\cos ^2\phi(t_i)
   }
\end{equation}
In other words, the discrete time evolution of the heading $\phi$ is given by iterating the $x$-dependent mapping

\begin{equation}
g(\phi)=\phi+ h k\frac{ 2 \sin \phi  (R+\cos \phi)-2 x \cos \phi}{\sin ^2\phi-\cos ^2\phi.
   }
\label{eq:jb:iterate}
\end{equation}
Differentiating, we obtain
\begin{equation}
g^{\prime}(\phi) = 1 +  \frac{2 h k (-2 -3 R \cos\phi +R \cos 3 \phi +3 x \sin \phi +x \sin3\phi
   )}{\cos ^2\phi -\sin ^2\phi }.
\label{eq:jb:quantized}
\end{equation}
The numerator is negative in the parameter range of interest, while the denominator is positive.  Hence, we can choose $k$ sufficiently small that $g$ is a contraction on $-\pi/4<\phi<\pi/4$ uniformly in $x$ in the range $-R<x<R$.  Thus the iterates of $\phi$ under the mapping (\ref{eq:jb:iterate}) converge to 0, and because $\dot x= -\sin\phi$, this proves the theorem. 
\end{proof}

\smallskip

\begin{remark}
Having established convergence to the desired equilibrium motion with this sample-and-hold Eulerian steering law, the next step is to show that the same path will be well approximated by a corresponding Lagrangian sample and hold result.  The final step is to show that interleaving very short straight line segments with the constant curvature law (\ref{eq:jb:sampled}) will also asymptotically align with the desired path.  As noted in \cite{Hildreth}, for constant velocity motions with no rotational component, heading direction can be determined accurately with short two to three frame sequences.
\end{remark}

The next step toward understanding path interpolation by alternating sequences of straight (for sensing) and curved (for steering) path segments is a heuristic model for calculating $\tau$ values along curved paths.  Specifically, given a feature point $(x_f,y_f)$, instead of computing $\tau=r/\dot r$ along a path $(x(t),y(t),\theta(t))$ of (\ref{eq:jb:BasicVehicle}), we compute the {\em quasi-linear time-to-transit}
\[
\tau^{*}(t)=r(x(t),y(t),\theta(t))/(\frac{\partial r}{\partial x}\dot x + \frac{\partial r}{\partial y}\dot y).
\]
The key is to produce a splined interpolation of the desired path by appropriately alternating between short straight path segments along which tau values for feature keypoints can be computed and curved segments whose constant curvatures are prescribed by tau values computed on the preceding straight segment. 

\smallskip

This value separates the component of image movement on the retina due to tangential speed along the path from the motion due to curving.  Space does not allow a complete exploration of steering laws based on $\tau^*$, but to summarize what's involved, the steps in comparing these with the Eulerian model are given in the algorithm below.

\begin{algorithm}
\caption{Determine steering signal from Lagrangian optical flow---{\em Sense - Perceive - Act}}
\begin{algorithmic}[1]
\item Choose a time interval $h$ that is compatible with the rate at which features enter and leave the field of view.
\item Form the partition: $t_0 < t_1 < \dots$ with $t_{j+1}-t_j = h$.
\item Define a segmentation protocol that groups feature images from each wall whose $\tau^*$ values and image locations determine the steering signal for the current time interval.
\item At each switching time $t_j$ return to 3 and repeat.
\end{algorithmic}
\end{algorithm}

\noindent {\bf Conclusion.}  The stylized models considered above do not capture all essential aspects of visual navigation, but they are capable of highlighting purely geometric components of factors that confound visual perception of motion.  Work is ongoing to understand how these factors are best dealt with in laboratory settings using state-of-the-art optical flow such as \cite{ilg2017flownet}.


\bibliography{references}
\bibliographystyle{IEEEtran}

\end{document}